# MOMENTUM COGGING AT THE FERMILAB BOOSTER*

K. Seiya#, C. Drennan, W. A. Pellico, K. Triplett, A. Waller, Fermilab, Batavia, IL 60510, USA

*Abstract*

The Fermilab Booster has an upgrade plan called the Proton Improvement Plan (PIP). [1] The flux throughput goal is 2E17 protons/hour which is almost double the current operation at 1.1E17 protons/hour. The beam loss in the machine is going to be an issue. The Booster accelerates beam from 400 MeV to 8GeV and extracts to the Main Injector (MI). The current cogging process synchronizes the extraction kicker gap to the MI by changing radial position of the beam during the cycle.[2][3] The gap creation occurs at about 700MeV which is about 6msec into the cycle. The cycle to cycle variations of the Booster are larger at lower energy. However, changing the radial position at low energy for cogging is limited because of aperture. Momentum cogging is able to move the gap creation earlier by using dipole correctors and radial position feedback, and control the revolution frequency and radial position at the same time. The new cogging is expected to reduce beam loss and not be limited by aperture. The progress of the momentum cogging system development is going to be discussed in this paper.

## PIP INTENSITY UPGRADE

The Booster accelerates beam from 400 MeV to 8GeV and extracts to the MI. The harmonic number of Booster is 84 and the rf frequency is 37.77MHz at injection. The 82 bunches are extracted to the MI at the rf frequency of 52.8114MH which is same as the MI injection frequency as shown in Figure 1(left).

The Booster creates two rf buckets length of extraction kicker gap by kicking two bunches out using notcher and knocker magnets during acceleration. The MI has a harmonic number of 588 and accepts a total of 12 batches from the Booster every 66.67 ms for the Nova operation which is scheduled in 2013. The batch from the Booster has to be injected to the specific MI bucket location.

In order to minimize beam loss in the Booster the gap creation should occur at as low an energy as possible. The PIP project has a goal of doubling the flux by doubling the number of cycles from the Booster. The increased cycle rate will double the beam loss in the notch creation process. Cogging is the process that synchronizes the extraction kicker gap to the MI injection bucket by controlling the revolution frequency.

A sweeping Booster rf frequency is compared to the MI fixed frequency and the difference of a bucket position between the Booster and the MI is accumulated during Booster cycle (Figure1, right). Frequency error from the injection time jitter and magnetic field error cause a variation of final bucket position at extraction.

The goal of momentum cogging is to create the gap at lower energy and to reduce orbit variations. This will reduce beam loss due to scraping and emittance growth due to beam orbit not centered in gradient magnets.

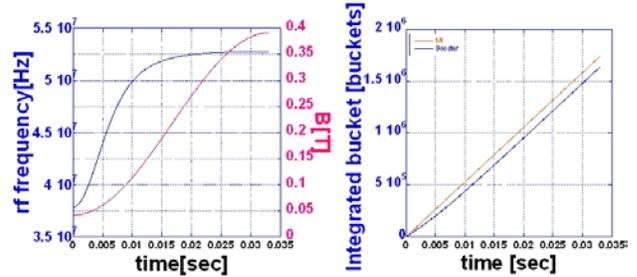

Figure 1: Rf frequency and magnetic field in the Booster cycle (left). The difference of a bucket position between the Booster and the MI (right).

## CURRENT COGGING OPERATION

Current cogging system controls revolution frequency: $f_{rev}$ by changing radial position: R of the beam during the cycle. The ratio between frequency and radial position changes is written in

$$\frac{\Delta f_{rev}}{f_{rev}} = (\frac{\gamma_t^2}{\gamma^2} - 1)\frac{\Delta(2\pi R)}{2\pi R}. \qquad (1)$$

The variation of the revolution frequency from cycle to cycle is larger at lower energy and it is hard to control by changing the radial position because of aperture limitations. The gap creation occurs at about 700 MeV which is 6 ms into the cycle and loses 2 bunches as shown in Figure 2. The radial position was changed by +/-3 mm before and after transition energy.

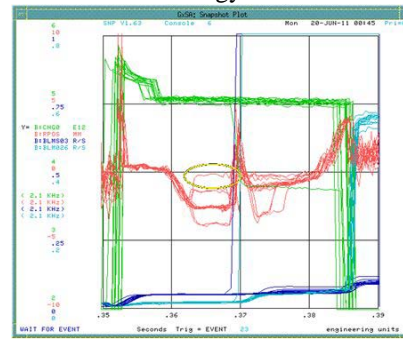

Figure 2: The beam intensity (green) and radial position (red) in a cycle.

## MOMNTUM COGGING WITH DIPOLE CORRECTOR

Momentum cogging is able to control the revolution frequency using the magnetic field from dipole correctors while a radial position feedback keeps the beam at the central orbit. The ratio between frequency and magnetic field changes under the fixed radial position is written in,

$$\frac{\Delta f_{rev}}{f_{rev}} = \frac{\Delta(2\pi R)}{2\pi R} - \frac{1}{\gamma^2}\frac{\Delta p}{p} = -\frac{1}{\gamma^2}\frac{\Delta B}{B}. \quad (2)$$

Forty eight correctors were installed in 2009[4]. A corrector has a length of 0.24 m and the dipole field strength is 0.009 T-m with 24.4 A. Maximum corrector current is +/- 40 A. A comparison between the Booster field at injection, 0.042 T, and the field from 48 dipole correctors with 10 A is calculated by

$$\frac{\Delta Bl_{corrector}}{B \times 2\pi R} = 0.0088, \quad (3)$$

when the Booster average radius is 75.75 m. The corrector should be able to compensate 1% field error at injection. The bucket slippage in 1 ms was calculated for current cogging and momentum cogging processes in Figure 3.

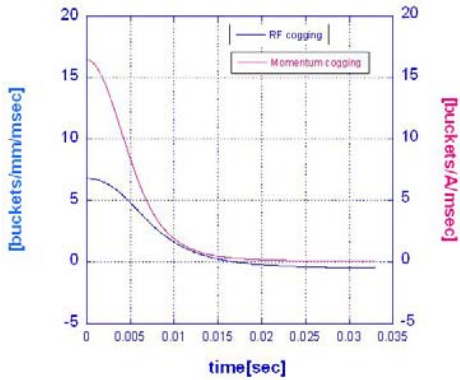

Figure 3: Bucket slippage in 1 ms: (blue) current cogging, (magenta) momentum cogging.

## MEASUREMENT WITH DIPOLE OFFSET

Beam studies have been performed using 24 correctors. A current offset of +4 A which is expected to make 95 buckets position change at 7ms was applied to the correctors from 2.5 ms to 7 ms (Figure 4). A gap was created at 2.3 ms. The bunch train was measured with a resistive wall monitor every 10 μsec and we counted the position from the gap. Five set of data were taken with and without 4A offset. The bucket position was changed by 90 buckets with 4 A offset in the measurements. (Figure 5) Beam position signals were taken at the same time and show that the orbit change with the offset was +/- 1mm (Figure 6).

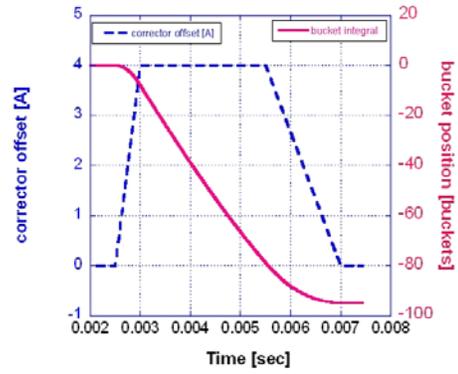

Figure 4: Corrector offset current (blue) and calculated bucket position (magenta).

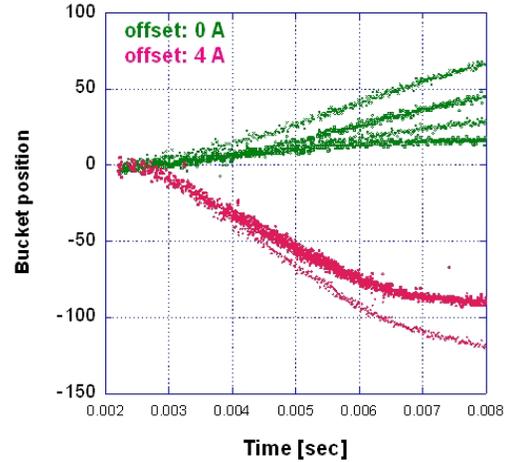

Figure 5: The bucket position with 4 A offset (magenta) and without offset (green).

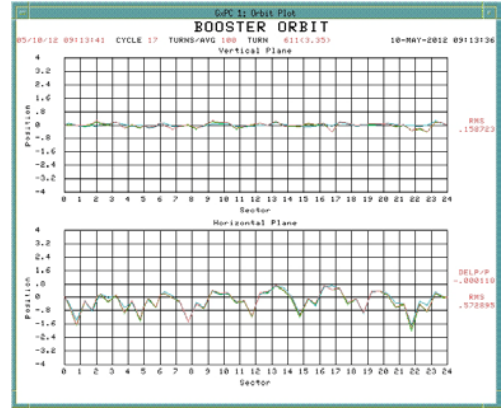

Figure 6: Differential orbits with 4A dipole offsets. Vertical scale is +/- 4mm.

## SUMMARY

Momentum cogging is a critical upgrade planed for the Booster. Momentum cogging is able to control the revolution frequency using the magnetic field from dipole correctors while a radial position feedback that keeps the beam at the central orbit. The idea has been successfully tested in beam studies and is expected to be operational in 2013.